\documentclass[12pt]{article}

  \usepackage{graphicx}
  \usepackage{epsfig}
  \usepackage{amssymb}
  \usepackage{subfigure}
  \usepackage{wasysym}
  \usepackage[T1]{fontenc}
  \usepackage{textcomp}

\evensidemargin - .5truecm
\begin{document}

\newcommand{\be}{\begin{equation}}
\newcommand{\ee}{\end{equation}}
\newcommand{\nn}{\nonumber}
\newcommand{\bea}{\begin{eqnarray}}
\newcommand{\eea}{\end{eqnarray}}
\newcommand{\bfig}{\begin{figure}}
\newcommand{\efig}{\end{figure}}
\newcommand{\bc}{\begin{center}}
\newcommand{\ec}{\end{center}}
\def\ad{\dot{\alpha}}
\def\ov{\overline}
\def\hlf{\frac{1}{2}}
\def\qrt{\frac{1}{4}}
\def\as{\alpha_s}
\def\at{\alpha_t}
\def\ab{\alpha_b}
\def\sq2{\sqrt{2}}
\newcommand{\smallz}{{\scriptscriptstyle Z}} %
\newcommand{\mz}{m_\smallz}
\newcommand{\smallw}{{\scriptscriptstyle W}}
\newcommand{\mw}{m_\smallw} 
\newcommand{\smallh}{{\scriptscriptstyle H}}
\newcommand{\mh}{m_\smallh}
\newcommand{\mt}{m_t}
\newcommand{\wh}{w_\smallh}
\def\th{t_\smallh}
\def\zh{z_\smallh}
\newcommand{\Mvariable}[1]{#1}
\newcommand{\Mfunction}[1]{#1}
\newcommand{\Muserfunction}[1]{#1}
%
%


\begin{titlepage}
\nopagebreak
{\flushright{
        \begin{minipage}{5cm}
         ROME1/1422/06\\
         IFIC/06-04\\
         RM3-TH/06-16\\
         IFUM-862/FT \\
        \end{minipage}        }

}
\renewcommand{\thefootnote}{\fnsymbol{footnote}}
\vskip 2.0cm
\begin{center}
\boldmath
{\Large\bf Two-loop electroweak corrections \\
\vspace*{2mm}
to Higgs production in proton-proton collisions}\unboldmath
\vskip 1.5cm
{\large  U.~Aglietti 
\footnote{Email: Ugo.Aglietti@roma1.infn.it} ,}
\vskip .2cm
{\it Dipartimento di Fisica, Universit\`a di Roma ``La Sapienza'' and
INFN, Sezione di Roma, P.le Aldo Moro~2, I-00185 Rome, Italy} 
\vskip .2cm
{\large  R.~Bonciani 
\footnote{Email: Roberto.Bonciani@ific.uv.es} ,}
\vskip .2cm
{\it 
Departament de Fisica Teorica IFIC, CSIC, Universitat de Valencia\\
Edifici d'Instituts de Paterna, Apt. Correus 22085, E-46071 Valencia, Spain
} 
\vskip .2cm
{\large G.~Degrassi\footnote{Email: degrassi@fis.uniroma3.it}},
\vskip .2cm
{\it Dipartimento di Fisica, Universit\`a di Roma Tre and 
INFN, Sezione di Roma III, \\ Via della Vasca Navale~84, I-00146 Rome, Italy} 
\vskip .2cm
{\large A.~Vicini\footnote{Email: Alessandro.Vicini@mi.infn.it}}
\vskip .2cm
{\it Dipartimento di Fisica, Universit\`a degli Studi di Milano and
INFN, Sezione di Milano,\\
Via Celoria 16, I--20133 Milano, Italy} 
\end{center}
\vskip 1.2cm

\begin{abstract}
We study the impact of the two-loop electroweak corrections
on the production of a Higgs boson via gluon-fusion in proton-proton
collisions at LHC energies.
We discuss the prescritpion to include the corrections to the hard
scattering matrix element in the calculation of the hadronic
cross-section $\sigma (p+p\to H+X)$.
Under the hypothesis of factorization of the electroweak corrections
with respect to the dominant soft and collinear QCD radiation,
we observe an increase of the total cross-section from 4 to 8\%, for
$\mh\leq160~{\rm GeV}$. This increase is comparable with the present 
QCD uncertainties originating from hard scattering matrix elements.

\end{abstract}
\vfill
\end{titlepage}    
\setcounter{footnote}{0}
\section{Introduction}

The Higgs boson is one of the missing ingredients of the Standard
Model and its discovery represents one of the most important physics 
goals of the LHC.
This goal will be achieved only if we can predict with high accuracy
all the production cross sections of this particle and if we understand 
in detail the different decay channels and the relative backgrounds.

At the LHC, the gluon-fusion is the dominant production mode over the entire 
range of interesting values of the mass of the Higgs particle ($100 \mbox{GeV} 
\apprle m_H \apprle 1 \mbox{TeV}$). In particular, in the range $100 \mbox{GeV} \apprle m_H 
\apprle 2 m_t$ this production mode is larger by almost one order of magnitude 
with respect to the next important mechanism, the vector boson fusion. It is,
therefore, very important to have a precise prediction of its cross section 
and a reliable estimate of the remaining theoretical accuracy.

The total cross section for the Higgs boson production by gluon fusion in the 
LO approximation was calculated in the late seventies \cite{H2gQCD0}. It is 
an ${\mathcal O}( \alpha_S^2 \, \alpha)$ 
calculation, since the Higgs couples to the 
gluons only via a heavy-quark loop (the most important contribution is the
one due to the loop of top). For what concerns the higher orders, the 
calculation of the NLO QCD corrections have 
been done in the infinite $m_t$ (mass of the top) limit in \cite{H2gQCD1},
and, with the full quark mass dependence, in \cite{QCDg2}.
Besides of the fact that the infinite $m_t$ approximation should be valid 
in the Higgs mass range $m_H \apprle 300$ GeV, it has been noticed 
\cite{Kramer:1996iq} that this approximation works also for values of $m_H$ 
beyond the top threshold, and up to masses of ${\mathcal O}(1 \mbox{TeV})$. 
The total effect of the NLO QCD corrections is the increase of the LO 
cross section by a factor 1.5--1.7, giving a residual 
renormalization/factorization scale dependence of about 30\%.
The electroweak corrections were evaluated in the 
infinite $m_t$ limit in \cite{GAMB} and turned out to amount to less
that 1\%.
The attention was driven by the evaluation of the NNLO QCD 
corrections, carried out in the infinite $m_t$ limit by several groups 
\cite{H2gQCD2}. The calculation shows a good convergence of the 
perturbative series: while the NNLO corrections are sizable, they are, 
nevertheless, smaller that the NLO ones. Moreover, the NNLO corrections 
improve the stability agaist renormalization/factorization scale 
variations. The effect due to the resummation of soft-gluon radiation has been 
included in \cite{bd4}, and the remaining theoretical uncertainty, due 
to higher-order QCD corrections, has been estimated to be smaller than 10\%.
Finally, several efforts were also devoted to the calculation of QCD radiative 
corrections to less inclusive quantities, such as the rapidity distribution,
recently evaluated at the NNLO \cite{Babis}, or the transverse momentum ($q_T$) 
distribution \cite{QT}, which, in \cite{Bozzi:2005wk}, is evaluated using the 
fixed-order perturbative results up to NLO in QCD and the 
resummation up to the NNLL.

Motivated by this accurate scenario, the NLO electroweak corrections to
the gluon fusion were reexamined recently. In \cite{ABDV} the  
contribution to the partonic cross section due to the light fermions were 
calculated. It turned out that they are sizeable. In particular, in the 
intermediate Higgs mass range, from 114 GeV up the the $2 m_W$ threshold, 
these corrections increase the LO partonic cross section by an amount of 
4--9\%. For larger values of the mass of the Higgs, $m_H > 2 m_W$,
they change sign and reduce the LO cross section; however, in this region 
the light-fermion corrections are quite small, reaching at most a $-2\%$. 
In \cite{DM}, also the remaining electroweak corrections due to the top quark 
were calculated as a Taylor expansion in $m_H^2/(4m_W^2)$. They are valid 
for $m_H \apprle 2 m_W$, range in which they have opposite sign with respect 
to the light-fermion corrections. However, the corrections due to the top quark
are smaller in size, reaching at most a 15\% of the light-quark ones.

The impact of the NLO electroweak corrections on the hadronic cross 
section has not been discussed yet. We present here the effect of their
inclusion in the calculation at the hadronic level.

\section{Inclusion of the Two-Loop Electroweak Corrections}

\begin{figure}
\begin{center}
\vskip-10.5cm
\hspace*{-2.5cm}
\includegraphics[height=280mm,angle=0]{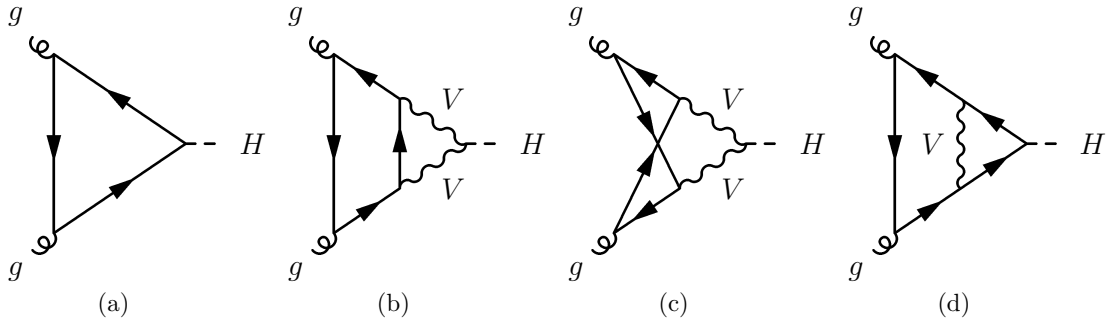}
\vskip-12cm
\caption{Lowest order (a) and generic NLO-EW (b), (c), (d) Feynman
  diagrams. The solid lines are fermions. The wavy lines are gauge
  bosons $(V=W,Z)$.}
\label{diagabc}
\end{center}
\end{figure}

The partonic gluon fusion process occurs, in lowest order, via one-loop
diagrams, as the one depicted in Fig.~\ref{diagabc} (a);
in the loop run only the top and the bottom quarks, because of the
Yukawa suppression of the lighter quarks.
The NLO-EW corrections are schematically represented by the diagrams
in Figs.~\ref{diagabc} (b), (c) and (d).
In particular, in Figs.~\ref{diagabc} (b) and (c) the WWH/ZZH couplings 
avoid the Yukawa suppression, and, therefore, in these diagrams the fermionic
line represents all the possible flavours: light flavours, evaluated in 
\cite{ABDV}, and top quark, evaluated in \cite{DM}.
In Fig.~\ref{diagabc} (d), instead, the fermionic line can represent only the
top quark \cite{DM}. 

At the hadronic level, we consider the Higgs boson production at the LHC,
and therefore in proton-proton collisions. The hadronic cross section can be 
written as:
\bea
&&\sigma(p+p\to H+X) =
\sum_{a,b}\int_0^1 dx_1 dx_2 \,\,f_{a,p}(x_1,M^2)\,
f_{b,p}(x_2,M^2)\times\nonumber\\
&&~~~~~~~~~~~~~~~~~~~~~~~~~~~\times
\int_0^1 dz~ \delta\left(z-\frac{\tau_H}{x_1 x_2} \right)
~\Big(1+\delta_{EW}(\mh)\Big)\hat\sigma_{ab}(z)\nonumber\\[2mm]
&&\hat\sigma_{ab}(z)=
\hat\sigma_0\,
\left(1+K^{QCD~only}_{ab}(\alpha_s(\mu^2),\mu^2,M^2)  \right)
\label{sigmafull}
\eea
where the partonic processes initiated by partons $(a,b)$ are
convoluted with the corresponding parton densities
$f_{i,p}(x,M^2),~~(i=a,b)$, evaluated at a scale $M$.
The effect of the higher order QCD and EW corrections is described by
the two functions $K^{QCD-only}$ and $\delta_{EW}$, obtained by factorizing 
the lowest order cross section $\hat\sigma_0$.

In the partonic cross section, QCD and EW corrections have been factorized.
This ansatz is valid up to subleading higher order corrections which
start at the 3-loop level 
(i.e. ${\cal O}(\alpha\alpha_s)$ with respect to the lowest order).
The factorization of the QCD initial state collinear divergences
holds for the hard process described by the electroweak NLO
corrections, following from general arguments of the factorization
theorems and from the universal nature of the initial state
collinear radiation.
In fact, the whole set of EW corrections
is characterized by a scale $\mw$, much harder than the one typical
of the leading collinear emission.
In addition, in the limit of light Higgs, the EW corrections can be
expanded as a Taylor series in powers of $\mh/\mw$ and the EW
corrections vertex becomes effectively pointlike.
In this regime the factorization of the QCD collinear divergences
becomes rigorous. 
For heavier Higgs masses, 
the factorization should still be valid only in leading order,
due to the modifications induced by the EW form factor.

\section{Numerical Results}

The hadronic proton-proton cross section has been calculated at LHC
energy, i.e. $\sqrt{S}=14$ TeV, 
in NNLO-QCD accuracy, i.e. setting $\delta_{EW}=0$,
using the MRST2002 NNLO parton distribution functions \cite{mrst2002}.
The theoretical uncertainty due to the choice of the renormalization
scale $\mu$ and of the factorization scale $M$ has been canonically estimated
by setting $M=\mu$ equal to $\mh/2,\mh,2\mh$ respectively.
The predictions, shown in Fig.~\ref{csLHC} (dotted lines),
vary by approximately $\pm 8$\% with respect to the central value.
This uncertainty is further reduced when including the effect of the
resummation of all the initial state soft gluon radiation \cite{bd4}.

\bfig
\bc
\begin{picture}(0,0)%
\includegraphics{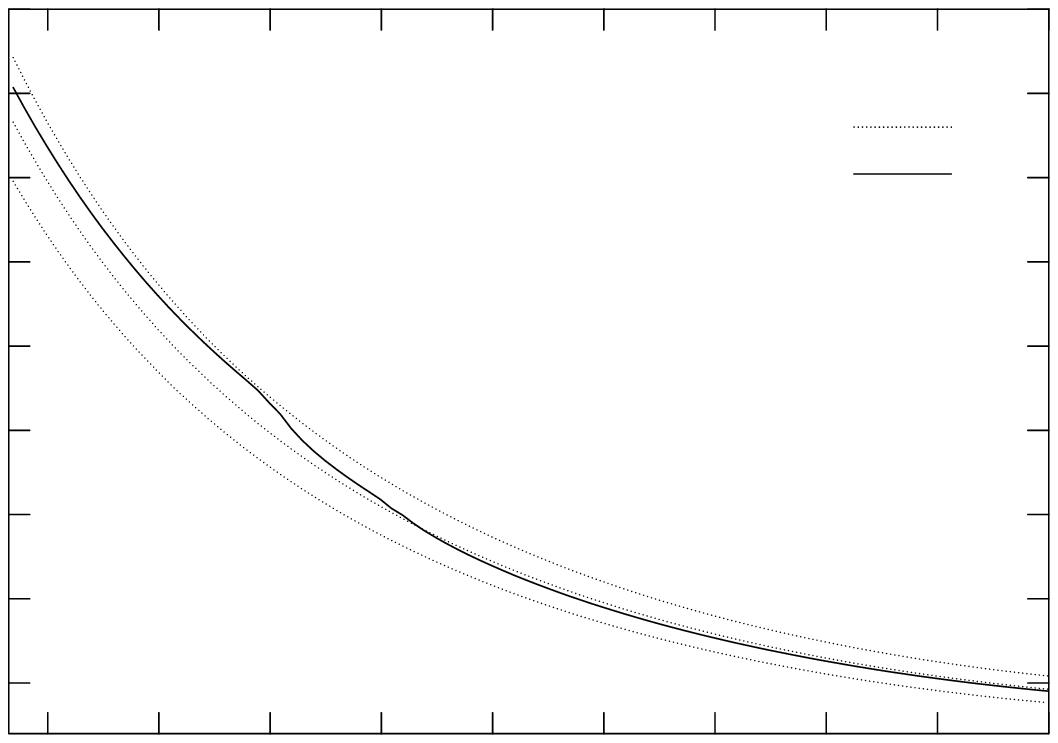}%
\end{picture}%
\setlength{\unitlength}{0.0200bp}%
\begin{picture}(18000,12960)(0,0)%
\put(1925,2707){\makebox(0,0)[r]{\strut{} 10}}%
\put(1925,3920){\makebox(0,0)[r]{\strut{} 15}}%
\put(1925,5133){\makebox(0,0)[r]{\strut{} 20}}%
\put(1925,6345){\makebox(0,0)[r]{\strut{} 25}}%
\put(1925,7558){\makebox(0,0)[r]{\strut{} 30}}%
\put(1925,8771){\makebox(0,0)[r]{\strut{} 35}}%
\put(1925,9984){\makebox(0,0)[r]{\strut{} 40}}%
\put(1925,11197){\makebox(0,0)[r]{\strut{} 45}}%
\put(1925,12410){\makebox(0,0)[r]{\strut{} 50}}%
\put(2761,1429){\makebox(0,0){\strut{} 120}}%
\put(4362,1429){\makebox(0,0){\strut{} 140}}%
\put(5964,1429){\makebox(0,0){\strut{} 160}}%
\put(7565,1429){\makebox(0,0){\strut{} 180}}%
\put(9167,1429){\makebox(0,0){\strut{} 200}}%
\put(10769,1429){\makebox(0,0){\strut{} 220}}%
\put(12370,1429){\makebox(0,0){\strut{} 240}}%
\put(13972,1429){\makebox(0,0){\strut{} 260}}%
\put(15573,1429){\makebox(0,0){\strut{} 280}}%
\put(17175,1429){\makebox(0,0){\strut{} 300}}%
\put(550,7029){\rotatebox{90}{\makebox(0,0){\strut{} $\sigma_{p+p \to H+X} (\mbox{pb})$ }}}%
\put(9687,275){\makebox(0,0){\strut{} $m_H (\mbox{GeV})$ }}%
\put(10769,6345){\makebox(0,0)[l]{\strut{}Upper line $\mu=0.5 m_H$}}%
\put(10769,5618){\makebox(0,0)[l]{\strut{}Central lines $\mu=m_H$}}%
\put(10769,4890){\makebox(0,0)[l]{\strut{}Lower line $\mu=2 m_H$}}%
\put(14097,10712){\makebox(0,0)[r]{\strut{}NNLO QCD}}%
\put(14097,10037){\makebox(0,0)[r]{\strut{}NNLO QCD + NLO EW}}%
\end{picture}
\caption{The cross section $\sigma_{p+p \to H+X}$, in pb, is plotted as a 
function of the mass of the Higgs boson, 
between 114 GeV and 300 GeV. 
The dotted lines describe the band of NNLO-QCD uncertainty, for three values
of the QCD factorization/renormalization scale $\mu=\mh/2,m_H,2\mh$.
The solid line is the NNLO-QCD $(\mu=\mh)$ with the two-loop EW corrections,
according to Eq.~(\ref{sigmafull}).
The two-loop EW corrections include also the top-quark effect, for
$\mh\leq 155$GeV, but only the light quarks contribution for larger
values of $\mh$. }
\label{csLHC}
\ec
\efig

The two-loop electroweak corrections have been added according to 
Eq.~(\ref{sigmafull}) and setting $M=\mu=\mh$. The light fermion corrections 
can be evaluated for any choice of $\mh$, whereas the top quark contribution 
has been computed by means of a Taylor expansion and is limited to the region 
$\mh\leq 160$ GeV. The hadronic cross section increases from 4 to 8\%, for 
$\mh\leq 160$ GeV. As we can observe in Fig.~\ref{csLHC}, the effect of the 
electroweak corrections is an increase of the cross section by an amount which 
is of the same order of magnitude of the NNLO-QCD theoretical uncertainty, and 
possibly larger than the uncertainty estimated after the resummation of soft 
gluon radiation. The main source of uncertainty on the hadronic cross section
remains in the accurate determination of the parton distribution functions of 
the proton.

The effect of the NLO-EW corrections is of great interest, 
because it enhances the most important Higgs production mechanism
and, in turn, affects the absolute number of events of all the Higgs
decay modes.

Following Eq.~(\ref{sigmafull}),
the NLO-EW corrections can be implemented as a simple rescaling
of the QCD hadronic cross section.
This multiplicative factor is presented in Table \ref{tabrescale} as a
function of $\mh$ and can be fitted, in the range $114 \, \mbox{GeV}
\apprle m_H \apprle 155 \, \mbox{GeV}$, by the following simple formula:
\be
\delta_{EW}(m_H) = 0.00961 + 6.9904 \cdot 10^{-5} \, \mh 
+ 2.31508 \cdot 10^{-6} \, \mh^2 \, .
\ee
\begin{table}[t]
\begin{center}
\begin{tabular}{|c|c||c|c||c|c||c|c|}
\hline
$\mh$ (GeV) & $\delta_{EW}$ & $\mh$ (GeV) & $\delta_{EW}$ 
& $\mh$ (GeV) & $\delta_{EW}$ & $\mh$ (GeV) & $\delta_{EW}$ \\
\hline
114 &  0.048 & 136 &  0.062 & 158 &  0.077 & 180 &  0.020\\
\hline
116 &  0.049 & 138 &  0.063 & 160 &  0.069 & 182 &  0.010\\
\hline
118 &  0.050 & 140 &  0.065 & 162 &  0.063 & 184 &  0.010\\
\hline
120 &  0.051 & 142 &  0.066 & 164 &  0.049 & 186 &  0.002\\
\hline
122 &  0.053 & 144 &  0.068 & 166 &  0.041 & 188 &  0.997\\
\hline
124 &  0.054 & 146 &  0.069 & 168 &  0.035 & 190 &  0.994\\
\hline
126 &  0.055 & 148 &  0.071 & 170 &  0.031 & 192 &  0.991\\
\hline
128 &  0.056 & 150 &  0.073 & 172 &  0.028 & 194 &  0.989\\
\hline
130 &  0.058 & 152 &  0.074 & 174 &  0.026 & 196 &  0.987\\
\hline
132 &  0.059 & 154 &  0.076 & 176 &  0.024 & 198 &  0.986\\
\hline
134 &  0.060 & 156 &  0.077 & 178 &  0.022 & 200 &  0.985 \\
\hline
\end{tabular}
\caption{Rescaling factor $\delta_{EW}$ as a function of the Higgs
  boson mass.}
\end{center}
\label{tabrescale}
\end{table}
The computation of the NLO-EW corrections to the gluon fusion process
has been described in detail in \cite{ABDV,DM}.
The analytical expression of the probability amplitude has been expressed in
terms of Generalized Harmonic PolyLogarithms (GHPL) \cite{ab} and
has been implemented in a FORTRAN routine\footnote{available upon request 
from the authors}.
The GHPL can be evaluated numerically in several different ways: by
direct numerical integration of the basic functions, by power
expansions or by solving the associated differential equations.
We have checked that these fully independent approaches agree.

\section{Conclusions}
In conclusion, the calculation of the QCD corrections to the production of a 
Higgs boson via gluon-fusion has reached a very high level of accuracy; 
the inclusion of the two-loop electroweak corrections, 
whose typical size for $\mh\leq 160$ GeV is larger than 5\% and then
comparable or larger than the QCD uncertainty, is highly desirable.
The main source of uncertainty on the hadronic cross section
remains in the accurate determination
of the parton distribution functions of the proton.

\vspace*{6mm}

\noindent {\bf Acknowledgments}

\noindent The authors wish to thank S. Catani, D. de Florian and M. Grazzini
for allowing the use of the numerical program 
of JHEP {\bf 0105} (2001) 025 [arXiv:hep-ph/0102227], and for useful 
discussions.

This work was partly supported by the European Union under the 
contract HPRN-CT2002-00311 (EURIDICE) and by MCYT (Spain) under 
Grant FPA2004-00996, by Generalitat Valenciana (Grants GRUPOS03/013 
and GV05/015).

\end{document}